\documentclass[twocolumn,showpacs,preprintnumbers,amsmath,amssymb]{revtex4}
\usepackage{dcolumn}% Align table columns on decimal point
\usepackage{bm}% bold math
\usepackage{graphicx}

\begin{document}

\title{Reconstructing the Neutron-Star Equation of State from Astrophysical
Measurements}

\author{Feryal \"Ozel}
\author {Dimitrios Psaltis}
\affiliation{University of Arizona, Department of Astronomy and Steward 
Observatory, 933 N. Cherry Ave., Tucson, AZ 85721}

\begin{abstract}
The properties of matter at ultra-high densities, low temperatures,
and with a significant asymmetry between protons and neutrons can be
studied exclusively through astrophysical observations of neutron
stars.  We show that measurements of the masses and radii of neutron
stars can lead to tight constraints on the pressure of matter at three
fiducial densities, from $1.85$ to $7.4$ times the density of nuclear
saturation, in a manner that is largely model-independent and that
captures the key characteristics of the equation of state.  We
demonstrate that observations with 10\% uncertainties of at least
three neutron stars can lead to measurements of the pressure at these
fiducial densities with an accuracy of 0.11~dex or $\simeq 30\%$.
Observations of three neutron stars with 5\% uncertainties are
sufficient to distinguish at a better than $3\sigma$ confidence level
between currently proposed equations of state.  In the electromagnetic
spectrum, such accurate measurements will become possible for
weakly-magnetic neutron stars during thermonuclear flashes and in
quiescence with future missions such as the International X-ray
Observatory (IXO).
\end{abstract}

\pacs{97.60.Jd, 26.60.Kp, 21.65.-f}

\maketitle

\section{Introduction}

Matter in the deep interiors of neutron stars is characterized by low
temperatures, large chemical potentials, and a significant asymmetry
between the number of neutron and protons. In these conditions,
several phenomena may take place, such as the creation of stable
bosons and hyperons or a phase transition to quark matter. In
addition, at these high densities, the symmetry energy of nucleonic
matter plays a dominant role in determining its microscopic
properties. The large uncertainty in understanding these phenomena has
led to a wide range of models for the equation of state of
neutron-star matter, which cannot be distinguished by hadron collision
experiments or with observations of the early universe \cite{Hands,
LP}.

The relation between the pressure $P$ and density $\rho$ of
neutron-star matter determines the macroscopic properties of the stars
and, in particular, their radii $R$, masses $M$, and moments of
inertia $I$. In fact, there is a unique map between the microscopic
$P-\rho$ relation and the macroscopic $M-R$
one~\cite{Lindblom}. Astrophysical observations of neutron stars aim
to exploit this mapping and invert it in order to measure their
equation of state.

The mass and radius of a neutron star of a given central density,
$\rho_{\rm c}$, depends on the entire relation between pressure and
density up to $\rho_{\rm c}$, because of the density gradient from the
center of the star to its surface. If the mass-radius relation could
be traced observationally with a large number of neutron stars
spanning the entire range of masses (e.g., 0.2--2~$M_\odot$), then the
mapping could be formally inverted to generate the complete equation
of state~\cite{Lindblom}. However, the astrophysical formation
channels for neutron stars limit them to masses larger than $\simeq
1.2 M_\odot$, rendering this approach impractical.

Prakash and Lattimer~\cite{LP} studied extensively the properties of
neutron-star models for a large range of equations of state and
concluded that the predicted radii depend primarily on the pressure at
a fiducial density, comparable to the nuclear saturation density of
$\rho_{\rm ns}\sim 2.7\times 10^{14}$~g~cm~$^{-3}$. More recently,
Read et al.~\cite{Read} showed that not only the approximate radii
of neutron stars, but a large number of observables depend on a small
number of parameters that characterize the equation of state.

In this paper, we show that the masses and radii of neutron stars act
as tracers of the pressure of neutron-star matter at three fiducial
densities. In other words, the complete mass-radius relation to high
accuracy can be reproduced for all proposed equations of state, when
the pressure at these three densities is specified. Therefore, a
one-to-one mapping exists between the pressure at the three densities
and the mass-radius relation, which can be inverted even if the
observed neutron stars are confined to a small range of masses. We
present the formalism for this inversion that takes into account the
uncertainties in the mass and radius measurements and show that
observations of three neutron stars with a 10\% uncertainty can lead
to a measurement of the pressure of ultra-dense matter at these
densities with comparable accuracy.

\section{A Minimal Representation of Neutron-Star Equations of State}

Inferring the pressure of neutron-star matter at a large number of
densities from astrophysical measurements would ideally generate the
equation of state in a model independent way. However, the range of
core densities of neutron stars with masses $\gtrsim 1.2 M_\odot$ is
in fact not very large and, therefore, the pressure at a small number
of fiducial densities $\rho_i$ governs their observable properties.
Choosing two fiducial densities, setting them to the fixed values
$\rho_1=1.85\rho_{\rm ns}$ and $\rho_2=2\rho_1$, and assuming a
piecewise polytropic relation between them has been shown to
accurately reproduce the complete pressure-density relation for a
large sample of proposed equations of state~\cite{Read}.

We explored the predicted mass-radius relations for the parametric
equation of state of Ref.~\cite{Read} in which the polytropic indices
between the fiducial densities are left as free parameters. We found
that different combinations of polytropic indices lead to practically
indistinguishable mass-radius relations for the neutron star models.
Consequently, this choice leads to a significant degeneracy between
the inferred uncertainties of the various model parameters when aiming
to reconstruct the equation of state from a set of mass-radius
measurements (see \S III).  This degeneracy is greatly reduced if we
choose, instead, the values of $P(\rho_1)$ and $P(\rho_2)$ to be the
free parameters and describe the equation of state at densities larger
than $\rho_2$ by a polytrope with normalization set by the pressure
$P(\rho_3)$ at a third fiducial density $\rho_3 > \rho_2$. Because the
density $\rho_3$ only serves to specify the polytropic index at
densities larger than $\rho_2$, its exact value does not affect the
final result. Hereafter, we set $rho_3 = 2 \times \rho_2$. Following
Ref.~\cite{Read}, we supplement this with the equation of state
SLy~\cite{Douchin} for the outer layers of the neutron star for
densities below a $\rho_0$, which we leave as a free parameter.

Specifically, we use hereafter the following parametric equation of
state of neutron-star matter, which depends on four parameters
$\rho_0$, $P_1\equiv P(\rho_1)$, $P_2\equiv P(\rho_2)$, and $P_3\equiv
P(\rho_3)$.  

\noindent {\em (i)\/} For $\rho\leq\rho_0$, we interpolate between the 
tabulated values of the SLy equation of state~\cite{Douchin}. We
define $P_0$ and $\epsilon_0$ to be the pressure and energy density at
$\rho_0$, respectively.

\noindent {\em (ii)\/} For $\rho_0<\rho\leq\rho_1$, we define 
\begin{equation}
\Gamma_1 \equiv \frac{\log(P_1/P_0)}{\log(\rho_1/\rho_0)}
\end{equation} 
so that the pressure in this density range is given by 
\begin{equation}
P = P_1 \left(\frac{\rho}{\rho_1}\right)^{\Gamma_1}
\end{equation}
and the energy density is 
\begin{equation}
\epsilon = (1+a_1)\rho + \frac{P_1}{\Gamma_1-1} \left(\frac{\rho}{\rho_1}
\right)^{\Gamma_1}, 
\end{equation}
where 
\begin{equation}
a_1 = \frac{\epsilon_0}{\rho_0}-1-\frac{P_1}{(\Gamma_1-1)\rho_0}
\left(\frac{\rho_0}{\rho_1}\right)^{\Gamma_1}. 
\end{equation}

\noindent {\em (iii)\/} For $\rho_1<\rho\leq\rho_2$, we set
\begin{equation}
\Gamma_2 \equiv \frac{\log(P_2/P_1)}{\log(\rho_2/\rho_1)}
\end{equation} 
so that the pressure in this density range is given by 
\begin{equation}
P = P_1 \left(\frac{\rho}{\rho_1}\right)^{\Gamma_2}
\end{equation}
and the energy density is 
\begin{equation}
\epsilon = (1+a_2)\rho + \frac{P_1}{\Gamma_2-1} \left(\frac{\rho}{\rho_1}
\right)^{\Gamma_2}, 
\end{equation}
where 
\begin{equation}
a_2 = a_1 + \frac{P_1}{(\Gamma_1-1)\rho_1} -
\frac{P_1}{(\Gamma_2-1)\rho_1}.
\end{equation}

\noindent {\em (iv)\/} For $\rho> \rho_2$, we set
\begin{equation}
\Gamma_3 \equiv \frac{\log(P_3/P_2)}{\log(\rho_3/\rho_2)}
\end{equation} 
so that the pressure in this density range is given by 
\begin{equation}
P = P_2 \left(\frac{\rho}{\rho_2}\right)^{\Gamma_3}
\end{equation}
and the energy density is 
\begin{equation}
\epsilon = (1+a_3)\rho + \frac{P_2}{\Gamma_3-1} \left(\frac{\rho}{\rho_2}
\right)^{\Gamma_3}, 
\end{equation}
where 
\begin{equation}
a_3 = a_2 + \frac{P_1}{(\Gamma_2-1)\rho_2}
\left(\frac{\rho_2}{\rho_1}\right)^{\Gamma_2} - 
\frac{P_2}{(\Gamma_3-1)\rho_2}.
\end{equation}

\begin{figure}
\centering
   \includegraphics[scale=0.43]{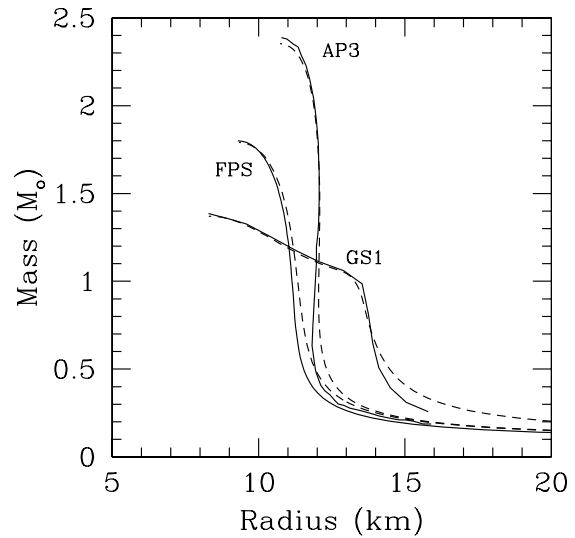}
\caption{Mass-radius relations for neutron stars, computed using the
complete (solid lines) as well as the parametric (dashed lines) forms
of three equations of state.}
\label{fig:M_R}
\end{figure}

\begin{table}[t]
\caption{\label{tab:table1}Best-fit EOS parameters for three sample 
equations of state. }
\begin{ruledtabular}
\begin{tabular}{lcccc}
EOS & $\log \rho_0$ & $\log P_1$ & $\log P_2$ & $\log P_3$ \\
\hline
FPS & 14.30 & 34.283 & 35.142 & 35.925 \\
GS1 & 14.85 & 34.504 & 34.884 & 35.613 \\
AP3 & 14.30 & 34.392 & 35.464 & 36.452 \\
\end{tabular}
\end{ruledtabular}
\label{table:EOS}
\end{table}

\begin{figure*}
\centering
   \includegraphics[scale=0.75]{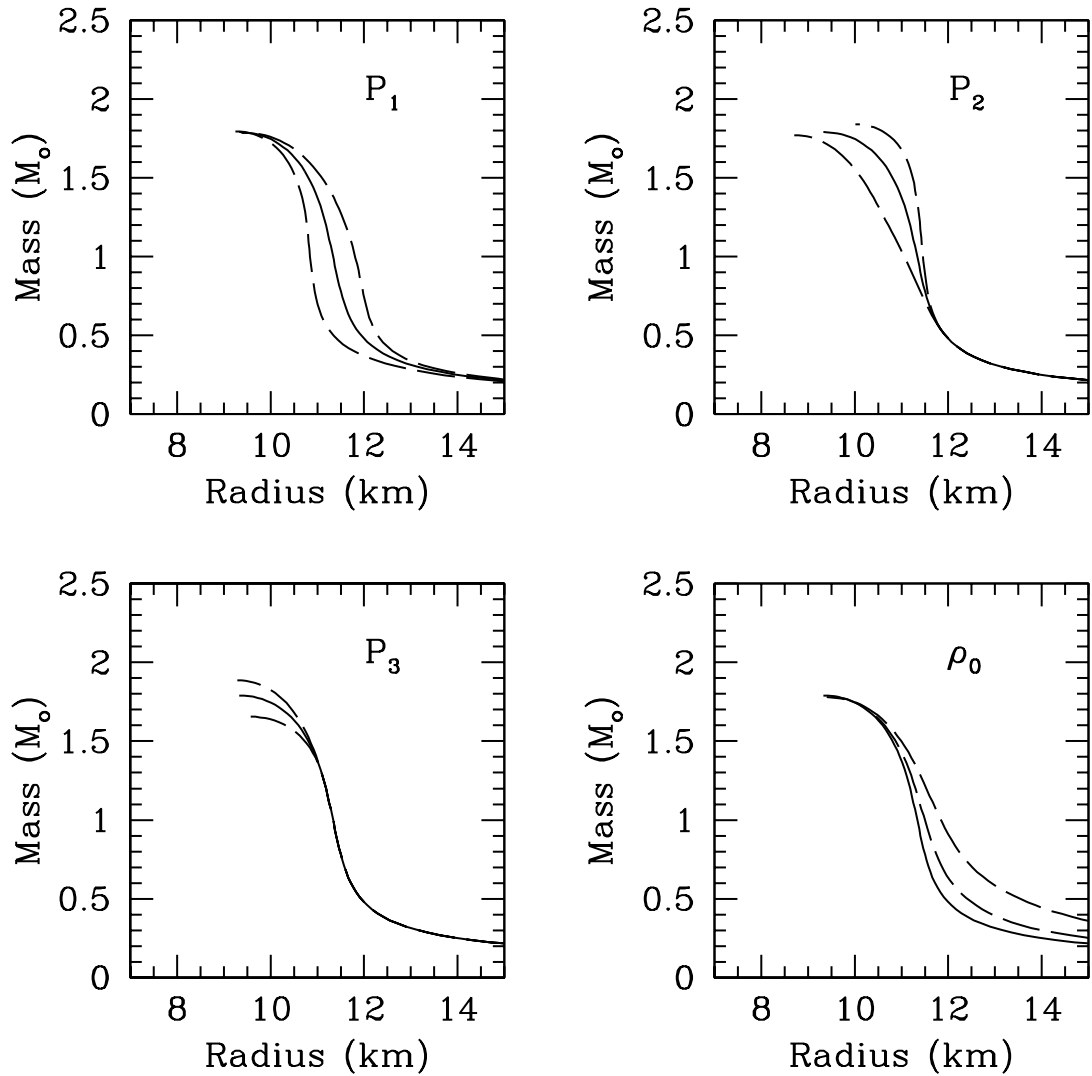}
\caption{The dependence of the calculated mass-radius relations of neutron
stars on the parametric equation of state. The first three panels show
the change in the predicted relation when the values of the parameters
$P_1$, $P_2$, and $P_3$ are varied by $25\%$ in each direction away from 
the best-fit values for the equation of state FPS. The last panel shows
the change in he predicted relation when the value of the parameter $\rho_0$
is reduced to 1/2 and 1/4 of its best-fit value for the same equation of
state.}
\label{fig:var_param}
\end{figure*}

Figure~\ref{fig:M_R} shows the mass-radius relations obtained by
integrating the Tolman-Oppenheimer-Volkoff equations with the complete
(solid lines) as well as the parametric (dashed lines) forms of a
sample of proposed equations of state, chosen to represent a wide
range of physical conditions (see Table~1 for the parameters). The
three examples also span a large range in neutron star masses and
radii. For neutron stars with masses $>1M_\odot$, the deviation in
radius between the models with the complete and parametric equations
of state is always $\leq 2$\%. The same is true for all the equations
of state considered in Ref.~\cite{Read}.

\begin{figure}
\centering
   \includegraphics[scale=0.43]{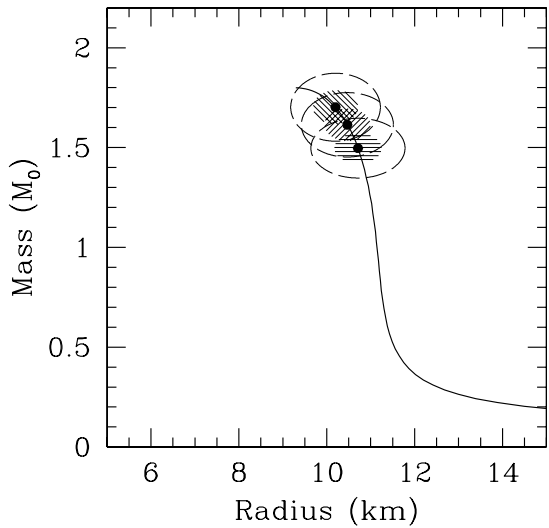}
\caption{Simulated values of the masses and radii of three neutron stars
that obey the FPS equation of state (solid dots). The hatch-filled
regions and the ellipses indicate 5\% and 10\% uncertainties in each
measurement, respectively.}
\label{fig:data}
\end{figure}

In order to get an intuitive understanding of the effect of these
parameters, we show in Figure~\ref{fig:var_param} how the calculated
mass-radius relations change when we vary each parameter in succession
away from the best-fit values for equation of state FPS.  The value of
the pressure at $\rho_1$ determines predominantly the characteristic
radii of neutron stars; this is the effect described in
Ref.~\cite{LP}. The value of the pressure at $\rho_2$ regulates the
slope of the mass-radius relation, at larger masses. The pressure at
$\rho_3$ sets the maximum mass of neutron stars. Finally, the value of
the density $\rho_0$, where the parametric branch of the equation of
state begins, only affects the mass-radius relation at small
neutron-star masses. In the first three cases, the parameters are
varied by 25\% away from their best fit value in each direction while,
in the last case, the density $\rho_0$ is reduced to 1/2 and 1/4 of
its original value in order to give rise to a visible difference in
the resulting mass-radius relation. Because of the much weaker
dependence of the observed properties of neutron stars on $\rho_0$,
especially at the mass range $\ge 1.2 M_\odot$ of interest, we fix
hereafter its value to $\log\rho_0=14.30$.

\begin{figure}
\centering
   \includegraphics[scale=0.43]{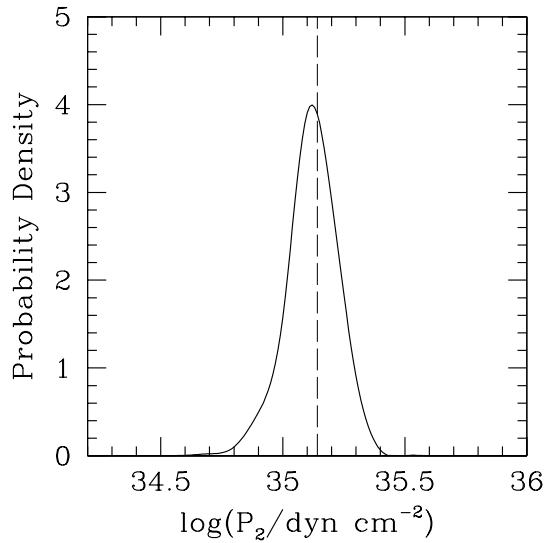}
\caption{The probability density over the pressure $P_2$ 
at 3.7 times the nuclear saturation density, evaluated when the other
two parameters of the equation of state are set to the best-fit values
for equation of state FPS. This is the probability that the parametric
equation of state reproduces the set of simulated data shown in
Figure~\ref{fig:data} that have 10\% measurement uncertainties. The
vertical dashed line indicates the true pressure at density $\rho_2$
of the FPS equation of state.}
\label{fig:param_P2}
\end{figure}

The distinct dependences of the mass-radius relations on the values of
the three parameters $P_1$, $P_2$, and $P_3$ shown in
Figure~\ref{fig:var_param} will allow a determination of each
parameter from observations with minimal correlated
uncertainties. Moreover, the rather strong dependence of the
mass-radius relations on the parameters of the equation of state
guarantees that these parameters can be accurately measured from
astrophysical observations. We discuss the expected uncertainties in
the following section.

\begin{figure*}
\centerline{
   \includegraphics[scale=0.43]{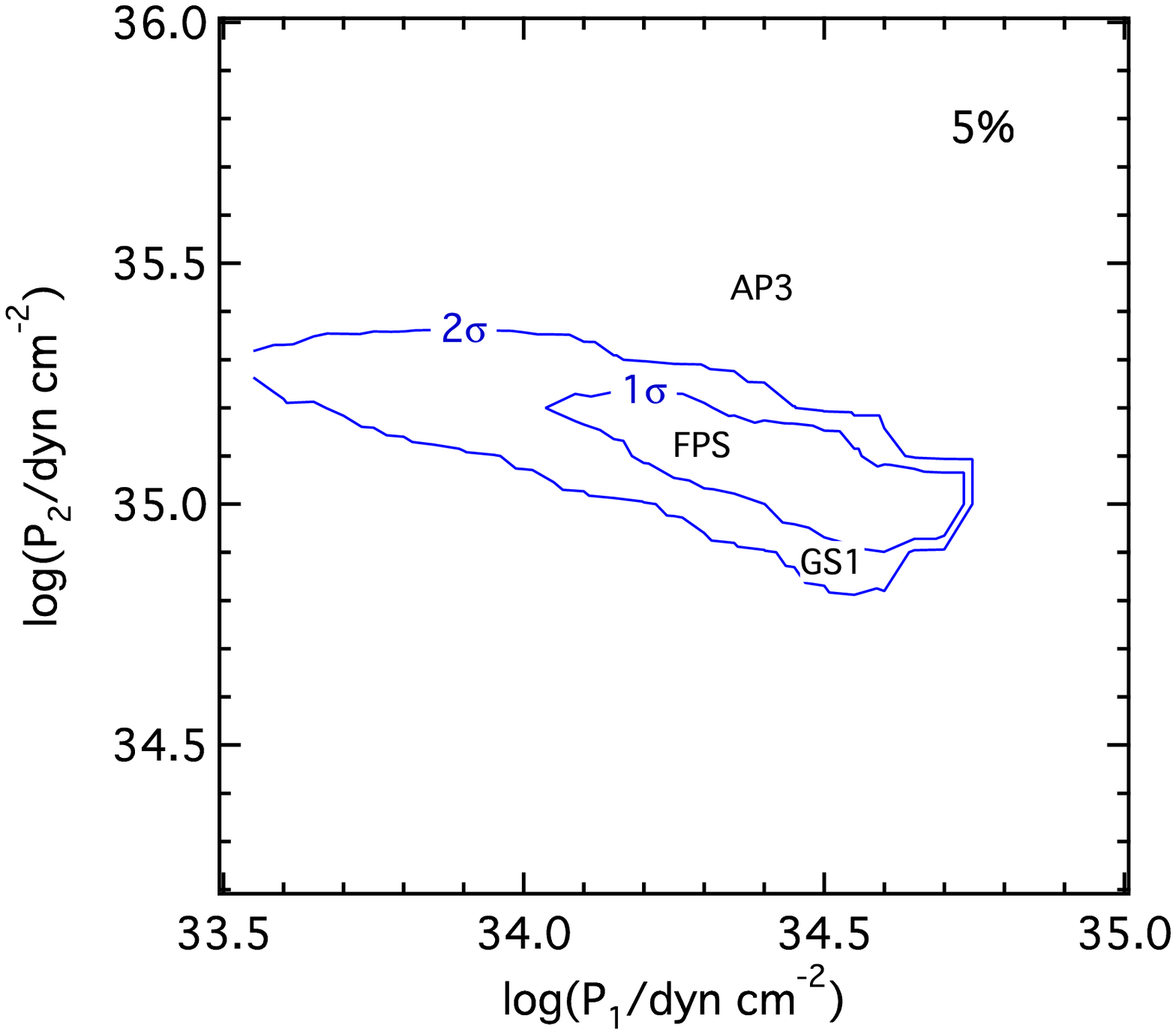}
   \includegraphics[scale=0.43]{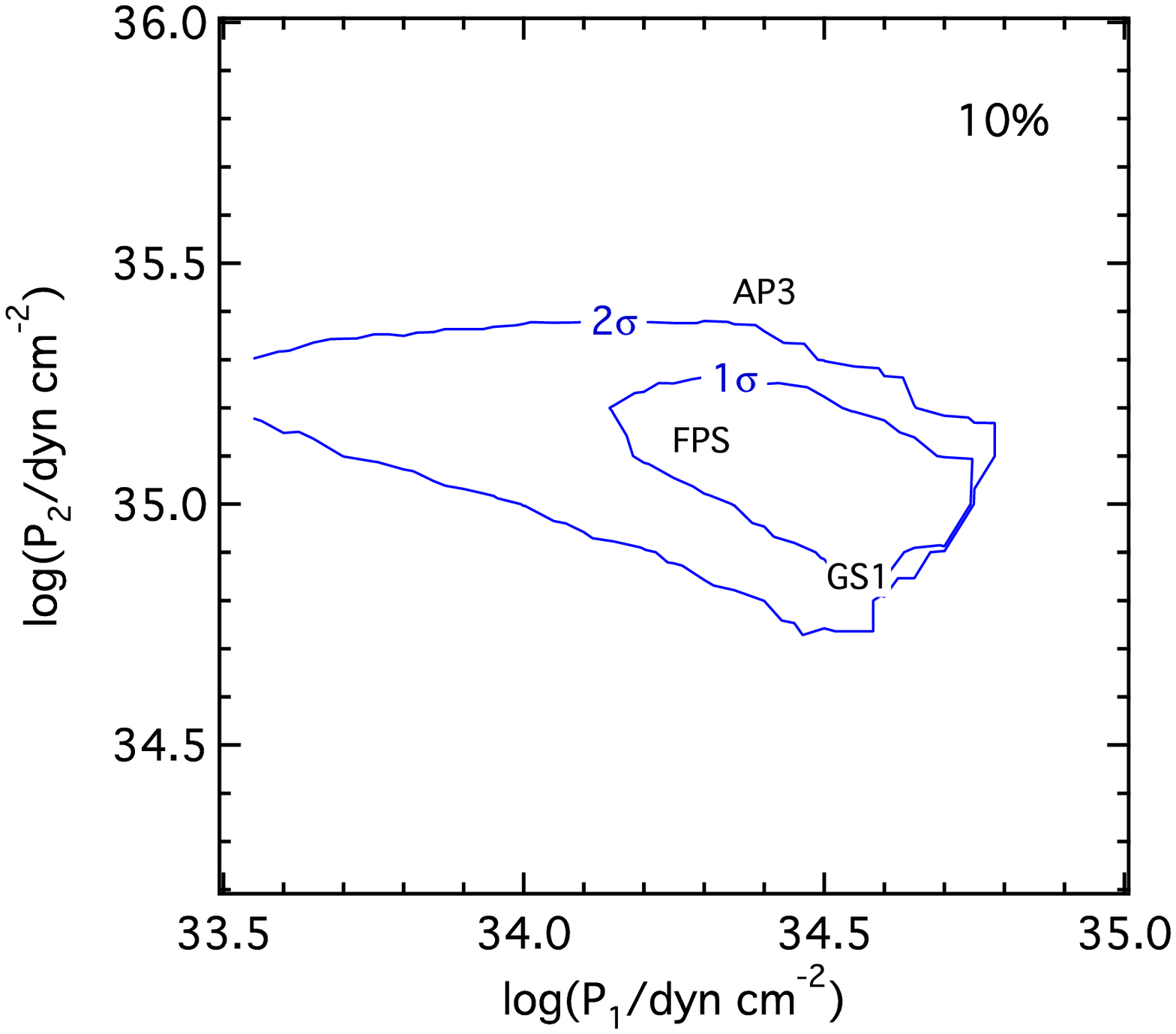}}
\centerline{
   \includegraphics[scale=0.43]{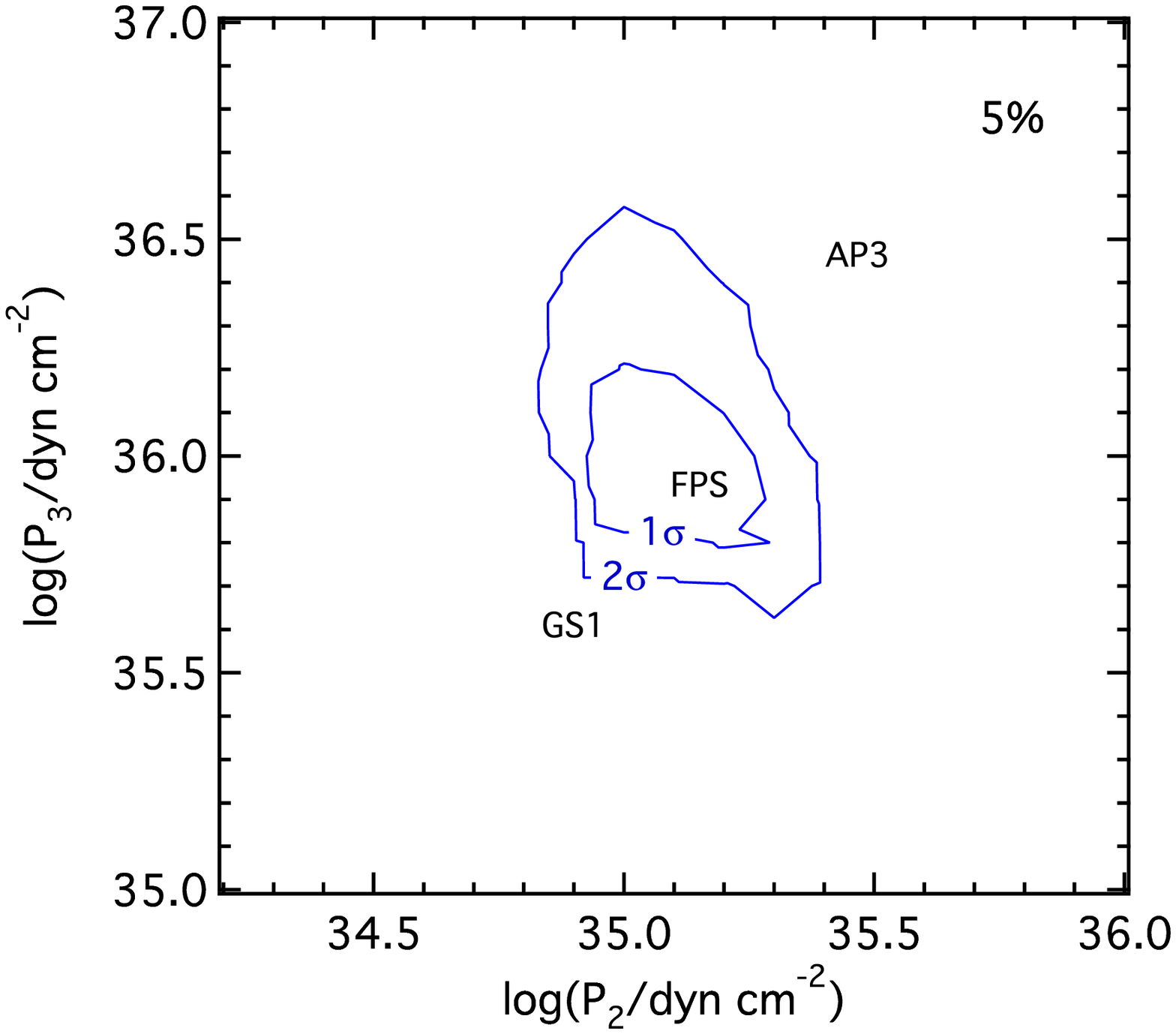}
   \includegraphics[scale=0.43]{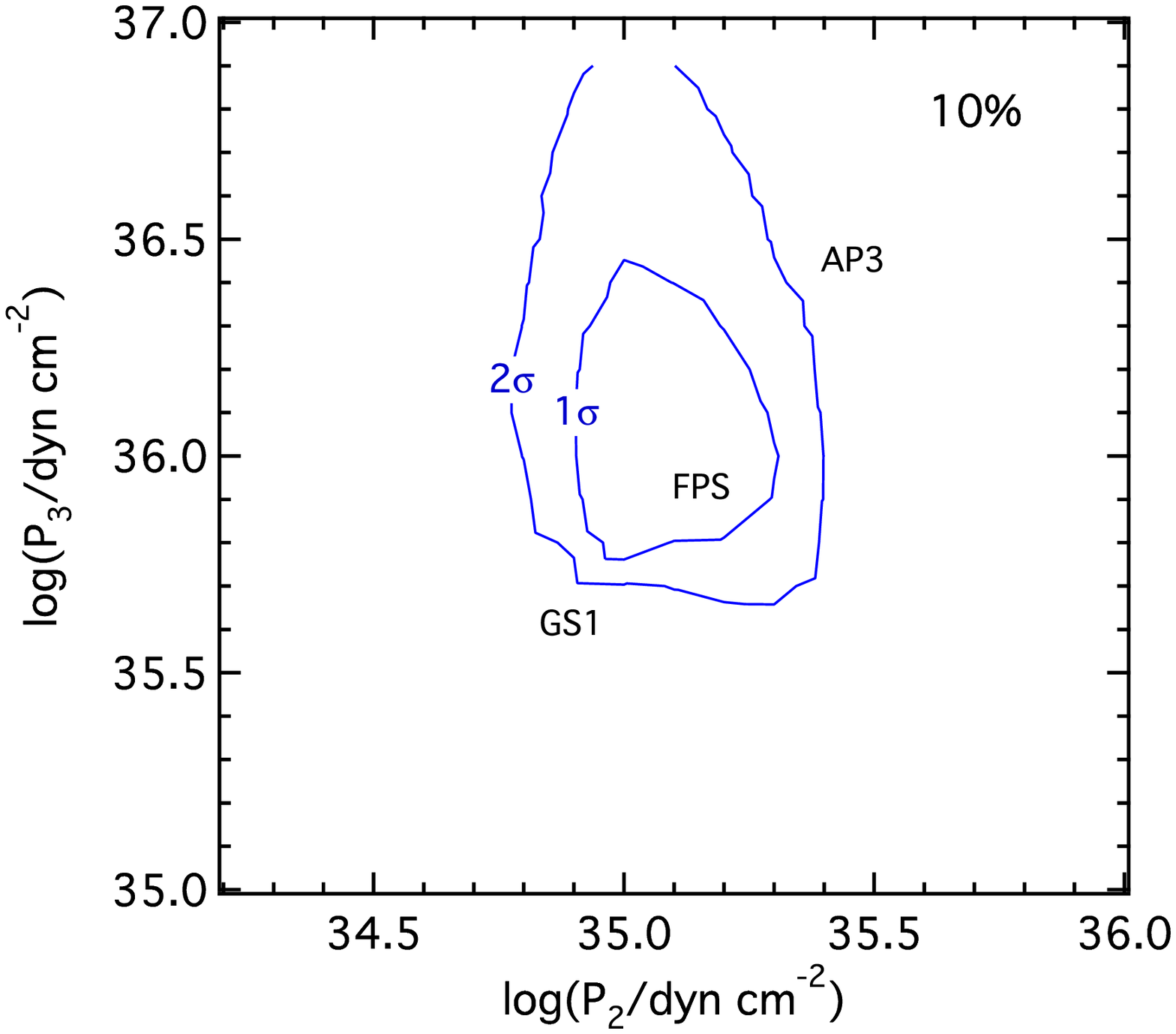}}
\caption{Contours of 68.2\% (1$\sigma$) and
95.4\% (2$\sigma$) confidence levels for the values of pairs of
parameters of the equation of state necessary to describe the
simulated data shown in Figure~\ref{fig:data}.  In each panel, the
probability density is marginalized over the parameter not shown. The
parameters $P_1$, $P_2$, and $P_3$ are the pressures at 1.85, 3.70,
and 7.40 times the nuclear saturation density, respectively. In all
panels, the corresponding true pressures for a sample of proposed
equations of state is shown for comparison. The left panels correspond
to a 5\% uncertainty in the measurements, whereas the right panels
correspond to a 10\% uncertainty. A 5\% measurement error is
sufficient to distinguish to a better than $3\sigma$ level not only
between equations of state that have different underlying physics but
also between ones with similar mass-radius predictions.}
\label{fig:param_2D}
\end{figure*}

\section{Equation of State Parameter Estimation from Measurements of Masses
and Radii}

The mass-radius relation of neutron stars that can be probed by
astrophysical observations is determined to a large extent by the
three parameters $P_1$, $P_2$, and $P_3$ of the parametric equation of
state that we discussed in the previous section. This is because the
value of the fourth parameter, $\rho_0$, only affects significantly
the properties of neutron stars with masses that are too low to be
astrophysically relevant.

For a given equation of state, the particular realization of a neutron
star is determined by its central density $\rho_{\rm c}$. As a result,
independent mass and radius measurements of $N$ neutron stars require
$N+3$ parameters to be completely modeled, under the assumption that
all stars follow the same equation of state. In other words,
observations of $N$ neutron stars yield $2N$ measurements, while
modeling their properties requires $N+3$ parameters, leaving
$2N-(N+3)=N-3$ degrees of freedom. Thus, in order to obtain a unique
solution, at least three neutron-star mass and radius measurements are
required. Hereafter, we discuss the situation with $N=3$.

Our aim is to invert the one-to-one map between the parameter space of 
the observables $(R_1, M_1, R_2, M_2, R_3, M_3)$ to the one of
the physical quantities $(P_1, P_2, P_3, \rho_{\rm c,1}, \rho_{\rm c,2},
\rho_{\rm c,3})$, where the indices on the masses, radii, and central
densities refer to the three observed neutron stars. Because of the
uncertainties inherent to the measurements, we actually need to
convert a probability distribution density over the parameter space of
observables ${\cal P}_{\rm obs}(R_1, M_1, R_2, M_2, R_3, M_3)$ to one
over the physical parameters ${\cal P}(P_1, P_2, P_3, \rho_{\rm c,1}, 
\rho_{\rm c,2}, \rho_{\rm c,3})$. The particular values of the
central densities for the three stars are not of particular interest
and obtaining the final probability density over the equation-of-state
parameters $P_1$, $P_2$, and $P_3$ would require marginalizing over
the three central densities. In order to avoid this additional
complication, we choose instead to transform the probability density
from the parameter space of observables to one over $(P_1, P_2, P_3,
M_1, M_2, M_3)$ and then marginalize over the three neutron star
masses. Formally, we write
\begin{eqnarray}
& &{\cal P}(P_1, P_2, P_3, M_1, M_2, M_3) =\nonumber\\
& &\qquad\qquad
{\cal P}_{\rm obs}(R_1, M_1, R_2, M_2, R_3, M_3)\nonumber\\
& &\qquad\qquad\qquad\qquad\qquad
\times{\cal J}\left(\frac{R_1,R_2,R_3}{P_1,P_2,P_3}\right)\;,
\label{eq:prob_transf}
\end{eqnarray}
where the last term is the Jacobian of the transformation between the
three observed radii and the three parameters of the equation of state.

For each set of parameters of the equation of state $P_1, P_2, P_3$,
we obtain the relation between neutron-star radius and mass, $R=R(M;
P_1, P_2, P_3)$, by integrating the Tolman-Oppenheimer-Volkoff equation
and use it to calculate numerically the Jacobian as
\begin{equation}
{\cal J}\left(\frac{R_1,R_2,R_3}{P_1,P_2,P_3}\right)\equiv\det
\left({\cal J}_{ij}\right)=\det
\left(\left.\frac{\partial R}{\partial P_i}\right\vert_{M_j}\right)\;.
\label{eq:jacobian}
\end{equation}
We then integrate equation~(\ref{eq:prob_transf}) over the three neutron star
masses to obtain the final probability distribution over the parameters
of the equation of state
\begin{eqnarray}
&&{\cal P}(P_1,P_2,P_3)=\int\int\int dM_1 dM_2 dM_3 \nonumber\\
&&\qquad\qquad\qquad
{\cal P}(P_1,P_2,P_3,M_1,M_2,M_3)\;.
\label{eq:inv_final}
\end{eqnarray}

In order to study the accuracy with which the equation of state
parameters can be inferred from astrophysical measurements and to
assess the potential correlations between them, we simulated
mass-radius data for three neutron stars that obey the FPS equation of
state, chosen here as an example. In particular, we set $M_1=1.50
M_\odot$, $R_1=10.7$~km, $M_2=1.61 M_\odot$, $R_2=10.5$~km, $M_3=1.70
M_\odot$, $R_1=10.2$~km, and assigned to each measurement an
independent Gaussian uncertainty with standard deviation equal to 5\%
and 10\% of the corresponding central values.  We show the simulated
data in Figure~\ref{fig:data} together with their uncertainties. Note
that this is a very conservative and narrow range of measurements for
neutron-star masses, as observed masses have been reported ranging
from $\simeq 1.2 M_\odot$ in radio pulsars~\cite{Thorsett99} to
$\simeq 2.1 M_\odot$ in accreting neutron stars~\cite{Ozel06}. We will
discuss at the end of this section the improved constraints on the
equation of state that can be obtained when a wider range of masses is
considered.

Using this set of simulated measurements, we inverted the mapping
between neutron star mass-radius and the equation-of-state pressures
at three fiducial densities and calculated the probability
distribution over the three parameters $P_1$, $P_2$, and $P_3$
according to equations~(\ref{eq:prob_transf}) to (\ref{eq:inv_final}).
Because the resulting probability density is defined in a three
dimensional parameter space, we show, in the following set of figures,
projections and marginalizations of this function along different
directions.

In Figure~\ref{fig:param_P2}, we show the normalized probability
density over the pressure $P_2$ at the fiducial density $\rho_2$,
evaluated when the other two parameters of the equation of state are
set to the best-fit values for equation of state FPS (see
Table~\ref{table:EOS}). The true value of the pressure $P_2$ for
equation of state FPS, shown by the vertical dashed line, is within
0.02~dex or 4.7\% of the peak of the probability distribution. Moreover,
the parameter $P_2$ is tightly constrained with an $1\sigma$ uncertainty of 
0.11 dex or 29\%.

In Figure~\ref{fig:param_2D}, we show the normalized two-dimensional
probability distributions over different pairs of the equation of
state parameters, marginalized over the remaining parameter. In each
panel, the two contours correspond to the 68.2\% (1$\sigma$) and
95.4\% (2$\sigma$) confidence levels. In all panels, the true
pressures for a set of representative equations of state are shown for
comparison. Finally, the left and right panels show the probability
densities obtained for the assumed 5\% and 10\% uncertainties in the
simulated data, respectively.

The results shown in the same figure provide additional justification
for the use of three (as opposed to fewer) fiducial densities at which
to measure the pressure of matter. In Ref.~\cite{Read}, which showed
that many proposed equations of state can be described by a piecewise
polytropic model, this was necessitated by that fact that the
potential production of bosons or hyperons at high densities or a
phase transition to quark matter causes abrupt changes to the equation
of state at multiple densities. Here, the use of three fiducial
densities allows us to distinguish between equations of state, without
requiring unattainable accuracy in the measurements.  In the examples
shown in Figure~\ref{fig:param_2D}, while both equations of state FPS
and GS1 are consistent with the simulated data in the $P_1-P_2$
parameter space, equation of state GS1 can be excluded at a confidence
better than $2-3\sigma$ in the $P_2-P_3$ space.

Of the three pressures $P_1$, $P_2$, and $P_3$, the one at 3.7 times
nuclear saturation density, i.e., $P_2$, is the most tightly
constrained by the data. This is because of the particular range of
masses that comprise our simulated data, for which the central density
is comparable to $\rho_2$. The pressure $P_1$ at 1.85 times nuclear
saturation density determines mostly the equation of state outside of
the cores of these neutrons stars and is, therefore, more tightly
constrained from above (see the top panels of
Figure~\ref{fig:param_2D}). On the other hand, the pressure $P_3$ at
7.4 times nuclear saturation density primarily affects the maximum
allowed mass for neutron stars and is, therefore, more tightly
constrained from below (see the bottom panels of
Figure~\ref{fig:param_2D}).

\begin{figure}
\centering
   \includegraphics[scale=0.43]{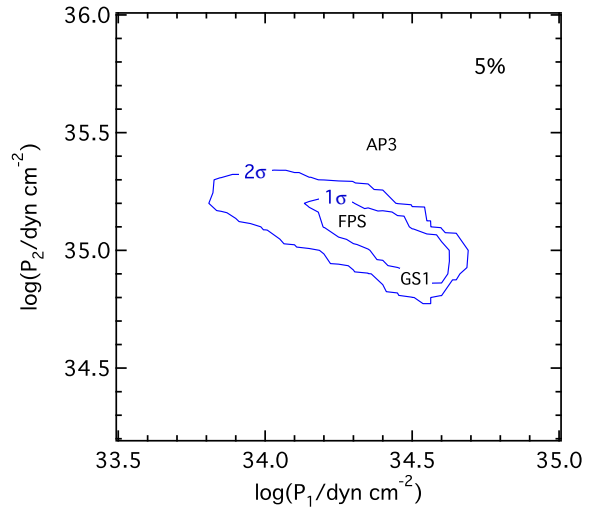}
\caption{The tightening of the constraints on the equation of state
parameter $P_1$ compared to the upper-left panel in
Figure~\ref{fig:param_2D}, when a $1.3~M_\odot$ neutron star is
included in the simulated data.}
\label{fig:better_mass}
\end{figure}

The above discussion suggests that allowing for a wider range of
masses in the simulated data would have improved the constraints on
$P_1$, if the range extended to lower neutron-star masses, and on
$P_3$, if the range included higher masses. We illustrate this for the
case of $P_1$ in Figure~\ref{fig:better_mass}, where we have assumed
that the three neutron stars have $M_1=1.31 M_\odot$, $R_1=10.95$~km,
$M_2=1.50 M_\odot$, $R_2=10.7$~km, and $M_3=1.70 M_\odot$,
$R_1=10.2$~km and allowed for 5\% measurement
uncertainties. Comparison with the top-left panel of
Figure~\ref{fig:param_2D} shows that the allowed range of values for
the parameter $P_1$ has shrunk significantly when a lower-mass neutron
star is included.

\begin{figure}
\centering
   \includegraphics[scale=0.43]{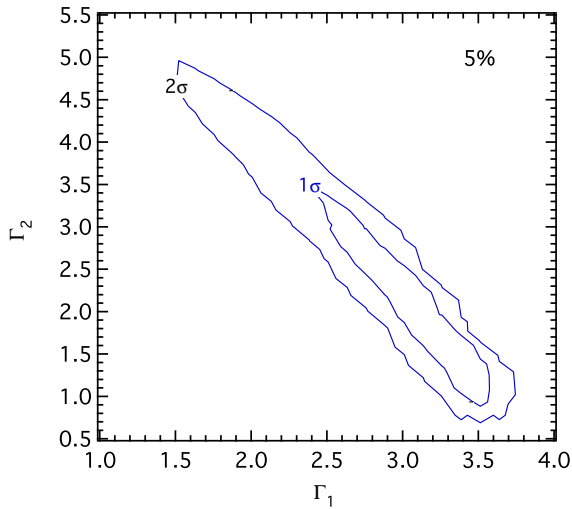}
\caption{The constraints on the polytropic indices $\Gamma_1$ and $\Gamma_2$, 
which appear in the alternative parametrization of the equation of
state, obtained for the same simulated data as in
Figure~\ref{fig:better_mass}.  This parametrization leads to
significantly correlated uncertainties even for the wide range of
neutron star masses considered here and for 5\% errors in the
measurements.}
\label{fig:gammas}
\end{figure}

It is evident from Figures~\ref{fig:param_2D} and
\ref{fig:better_mass} that the uncertainties in the three pressures
$P_1$, $P_2$, and $P_3$ are weakly correlated. In fact, there is no
discernible correlation between $P_3$ and the other two parameters.
This motivated our choice of the three pressures as the parameters of
the minimal representation of the equation of state as opposed to the
polytropic indices between the corresponding
densities~(cf.~\cite{Read}). Using the parametrization with polytropic
indices would have resulted in correlated uncertainties, such as those
shown in Figure~\ref{fig:gammas} for the case of $\Gamma_1$ and
$\Gamma_2$, even for the most constraining case of neutron star masses
and measurement errors considered above.

\section{Discussion}

In this paper, we have shown that measuring the masses and radii of
three neutron stars with a 10\% uncertainty can constrain to similar
accuracy the pressure of neutron-star matter at densities several
times the density of nuclear saturation. Moreover, astrophysical
measurements with 5\% uncertainties can distinguish, to a better than
$3\sigma$ confidence level, not only between equations of state that
have different underlying physics (e.g., FPS and GS1 in
Figure~\ref{fig:param_2D}) but also between ones with very similar
predicted mass-radius relations (e.g., FPS and AP3;
see~Figure~\ref{fig:M_R}).

Several types of observations in the electromagnetic spectrum have
already led to a number of mass and radius measurements for neutron
stars in a wide variety of astrophysical systems. Dynamical
measurements of the masses of radio pulsars in binary systems have
uncertainties as low as 0.1\%~\cite{Thorsett99}, but do not provide
any information on the radii of the stars. As such, these observations
can be used only to set lower bounds on the pressure at the three
fiducial densities (and especially at $\rho_2$ and
$\rho_3$)~\cite{Read}. Similar constraints, albeit with much larger
uncertainties, can also be set by the dynamical measurements of
neutron-star masses in X-ray binaries, such as those reported for
Vela~X-1~\cite{Barziv} or Cyg~X-2~\cite{Orosz}.  At the same time, a
secure identification as a black hole of a compact object that has a
dynamically measured mass of $\simeq 2 M_\odot$ would place a
stringent upper limit on the same parameters.

Measurements of both the masses and radii of neutron stars have been
reported for weakly magnetic neutron stars in low-mass X-ray binary
systems. In a number of cases, measurements were performed for globular
cluster X-ray transients in quiescence, when the accretion practically
ceases and the surface emission of the cooling neutron star can be
directly observed~\cite{heinke,webb}. These observations constrain the
apparent radii of neutron stars and, therefore, lead to measurements
of their masses and true radii with significant degeneracies. The
degeneracies can be overcome, however, if the masses of the neutron
stars are measured independently with dynamical observations of the
binary systems in which they reside. Current observations with
XMM-Newton led to $\simeq 15$\% uncertainties in the measurements of
the apparent radii~\cite{webb}. Using a telescope with an effective
area of 2~m$^2$, such as the one proposed for the International X-ray
Observatory, would reduce this uncertainty to $\sim 5$\%.

A different set of measurements of both the masses and radii of weakly
magnetic neutron stars were recently reported that utilized
observations of their thermal emission during thermonuclear
bursts~\cite{Ozel06,OzelGuver}. The degeneracies between the inferred
masses and radii were overcome in this case because up to three
distinct observables that have different dependencies on the masses
and radii were used. In this case, the measurements were performed
with the Rossi X-ray Timing Explorer and resulted in uncertainties of
$\simeq 10$\%. Reducing the uncertainties to 5\% requires an increase
of the effective area of the telescope by a factor of $\simeq 4$,
i.e., an effective area of $3.2$~m$^2$. This is well within the
design specifications of the Advanced X-ray Timing Array (AXTAR).

The masses and radii of neutron stars can in principle be constrained
also by observing and modeling in detail the flux oscillations at the
spin frequency of the neutron stars during the spreading of
thermonuclear flashes~\cite{Strohmayer}. This is not currently
possible because it requires count rates significant larger than those
achieved by the Rossi Explorer even for the brightest sources. Such
measurements, however, may be achieved with future observations.

Dynamical measurements of the masses and moments of inertia of at
least three neutron stars may also lead to very similar constraints as
those discussed in this work~\cite{inertia,Read}. Indeed, the moments
of intertia of neutron stars of known masses encode not only
information about their radii but of their density profiles, as
well. As such, they are sensitive to the parameters of the equation of
state at low densities and will lead to improved constraints on the
parameters $P_1$ and $\rho_0$. For the case of the double pulsar, a
measurement of the moment of inertia with an uncertainty of order 10\%
will probably be achieved by 2020~\cite{kw09}.

Finally, observations of the end stages of the coalescence of double
neutron stars with gravitational wave observatories such as LIGO,
VIRGO, and GEO-600 will open new avenues for measuring the masses and
radii of neutron stars and, hence, for determining the properties of
the equation of state of ultradense matter~\cite{ReadGW}.

\begin{acknowledgments}
We thank Tolga G\"uver for many useful discussions on the
astrophysical measurements of neutron-star masses and radii, as well
as Jocelyn Read and John Friedman for correspondence on the parametric
equation of state. We also thank Michael Kramer for useful discussions
as well as Greg Cook and Jim Lattimer for providing us with tables of
proposed equations of state. F.\,\"O.\ and D.\,P.\ are supported by
the NSF.
\end{acknowledgments}

\bibliographystyle{apsrev}

\end{document}